\documentstyle[prd,aps,preprint]{revtex}
        		
\newcommand\sk{{_{\bf k}}}

\newcommand{\sub}[1]{_{\mbox{\scriptsize#1}}}
\newcommand{\su}[1]{^{\mbox{\scriptsize#1}}}

\newcommand\ee{\end{equation}}
\newcommand\be{\begin{equation}}
\newcommand\eea{\end{eqnarray}}
\newcommand\bea{\begin{eqnarray}}

\newcommand\GeV{\,\mbox{GeV}}
\newcommand\MeV{\,\mbox{MeV}}

\newcommand\eV{\,\mbox{eV}}

\newcommand\mone{^{-1}}

\newcommand\mpl{M_{Pl}}

\newcommand\lsim{\mathrel{\rlap{\lower4pt\hbox{\hskip1pt$\sim$}}
    \raise1pt\hbox{$<$}}}
\newcommand\gsim{\mathrel{\rlap{\lower4pt\hbox{\hskip1pt$\sim$}}
    \raise1pt\hbox{$>$}}}

\newcommand\diff{\mbox d}

\begin{document}

\preprint{LANCS-TH/9615, hep-ph/9609441}
\draft
\tighten

\title{Cosmological consequences of particle creation during inflation}
\author{David H.~Lyth and David Roberts\\
School of Physics and Materials, University of Lancaster\\
Lancaster LA1 4YB,~~~U.~K.}
\date{September 1996}
\maketitle
\begin{abstract}
Particle creation during inflation is considered.
It could be important for species whose interaction 
is of gravitational strength or weaker.
A complete but economical formalism is given for 
spin-zero and spin-half particles,
and the particle abundance is estimated on the assumption that
the particle mass in the early universe 
is of order the Hubble parameter $H$. It is roughly the same for both 
spins, and it is argued that the same estimate should hold for higher 
spin particles in particular the gravitino.
The abundance is bigger than that from the usual 
particle collision mechanism if the inflationary energy scale
is of order $10^{16}\GeV$, but not if it is much lower.
\end{abstract}

\section{Introduction}

According to present ideas, particles species can be created 
in the very early universe by a number of 
mechanisms \cite{KT,Linde}.
The most familiar, which applies to particles of any spin, is the 
collision or decay of other particles. A number of other reasonably 
familiar mechanisms come into play for the case of a spin zero particle,
which corresponds to the oscillation of a scalar field.
If the field is initially displaced from its vacuum value
it will oscillate homogeneously,
a familiar example being the oscillation of the
inflaton field after inflation. The homogeneous oscillation can in turn 
cause further particle production, either through the decay
of the initial particles as in the original discussion of reheating 
after inflation, or through collective effects as in the more modern
discussions of `preheating'. Alternatively, the scalar field can be 
radiated from topological defects, as in the example of axion production 
from axionic strings.

The present paper explores a mechanism which is different from any of 
the above, namely the production of 
particles from the vacuum fluctuation during inflation.
We estimate the abundance of
particles arising from 
the vacuum fluctuation during inflation, and compare it
with the abundance from the particle collision mechanism
and with the cosmological constraints.

In contrast with earlier work, we take seriously the fact that
the effective mass $m(t)$ in the early universe is not expected to be
the same as the
true mass $m$ which is relevant at the present day, because
of the effect of interactions.
On the usual assumption that we are dealing with a supersymmetric
theory, it has long been recognised that even an interaction of 
gravitational strength will typically give a spin zero particle
a mass of order $H$ during inflation 
if $H$ is bigger than the true mass \cite{dfn,Ross,fvi}. 
More recently it has been emphasised 
\cite{Dine} that the same result is to be expected
even after inflation.
Although it has not so far been stated 
explicitly in the literature, one may expect this result to apply equally 
to particles of any spin.

This contribution of order $H$ to the mass
can be evaded if the mass is protected by
 a symmetry. For example
gauge bosons cannot acquire mass unless the symmetry is spontaneously 
broken, and Peccei-Quinn symmetry should keep the axion mass small,
but neither of these cases is of interest in the present
context (gauge bosons are too short-lived and the axion density 
will be dominated by other production mechanisms).

During inflation there are more possibilities for evading the mass
of order $H$. A number are available for spin zero, as reviewed for 
example in \cite{mynew} for the 
particular case of the inflaton field. Another case is that 
of the gravitino; its mass is proportional to 
$|W|$ where $W$ is the superpotential, and in some models of inflation
\cite{fvi,ewanmodels} it is much less than $H$ during inflation.
However, even if the contribution of order $H$ to the 
mass is absent during inflation it still 
typically reappears afterwards.

 If the interaction  is stronger 
than gravitational strength, one 
presumably expects a bigger mass \cite{ewanpersonal}, but  
in the present context this case is not of interest
because it will typically be associated with production mechanisms
that dominate the vacuum fluctuation. 
In particular, if the interaction is strong enough
to maintain thermal equilibrium the mass is of order the 
temperature $T$, but then the thermal abundance dominates.

We shall assume that $m(t)$ is not varying rapidly on the timescale of
the expansion rate $H$. Rapid variation will occur if $m(t)$
is generated by a coupling to a homogeneous oscillating scalar field,
because the oscillation frequency is necessarily big compared with $H$
(and so usually is  the amplitude). But in that case we are 
dealing with a separate mechanism of particle creation
\cite{paramres,dolkir,dolfreese}, which 
is decoupled from the vacuum fluctuation that we consider here.

With the above discussion in mind we shall assume
that a species with mass $m$ has an effective mass
$m(t) =\mu H$ in the early universe, with $\mu$ a constant of order
1. This should give an order of magnitude estimate, though
$\eta$ may actually be expected to have different values
during and after inflation.

The layout of this paper is as follows.
In Section 2 we give the
formalism for calculating the occupation number per quantum state,
treating separately the cases of spin $0$ and spin $1/2$.
In Section 3 we estimate the corresponding cosmological
abundance and compare it with the abundance coming from
particle collisions, on the assumption that these are of only 
gravitational strength. In Section 4 we summarize the 
conclusions, and some specific cases are worked out in an Appendix.

\section{Particle creation}

In this section we give the basic formalism for the creation of spin
zero and spin half particles. The reader who is interested only in the 
results  may skip to the next section, where the rather 
simple conclusion is summarized.

In considering particle creation from the vacuum fluctuation,
we are interested only in species which have very weak 
interactions. One reason is that a species with significant
interactions will achieve thermal 
equilibrium, losing all memory of its earlier 
abundance. The other is that there are strong cosmological 
constraints on the abundance of a long-lived species, whereas
short-lived species with stronger interactions decay before they can 
have any direct cosmological effect. 

In the case of a scalar field 
the quantum fluctuation is competing with a possible
classical displacement of the field from its vacuum value,
and this latter effect must be absent if the quantum fluctuation
is to be significant. If the moduli fields of superstring theory exist
with masses of order $100\GeV$, a 
displacement for them at the classical level
leads to a gross over-abundance in the standard cosmology,
which is called the `moduli problem' 
\cite{problem,dilaton,Banks,Steinhardt,Randall,Dine,gutti,thermal}.
As far as one can tell, the only way of 
solving the problem seems to be to dilute the moduli abundance with
a late bout of
`thermal' inflation as discussed in detail in Refs.~%
\cite{thermal,mythermal,ewanthermal,kimthermal}
(for earlier incomplete discussions of thermal inflation see
the works cited in Ref.~\cite{gutti}). 
However, this would also dilute the 
abundance of particles produced by the vacuum fluctuation 
that we are considering. (The vacuum fluctuation during thermal 
inflation is likely to be insignificant since the particle masses are 
likely to be much bigger than $H$, which is much less
than $100\GeV$.) 
Thus, for our discussion to be relevant in the context of supergravity
the moduli problem has to be eliminated by requiring a mass 
$\gg 100\GeV$, or else no classical 
displacement, rather than by modifying the standard cosmology.

As we are dealing with the early universe we can assume that the density 
is critical, so that the 
line element can be written 
\be
ds^2=a^2(\eta)[d\eta^2-dx^2-dy^2-dz^2]
\label{metric}
\ee
where $a$ is the scale factor and
$\eta$ is conformal time, related to the proper time of a comoving 
observer by $d\eta=dt/a$. Using the time coordinate 
$x^0\equiv \eta$
and comoving space coordinates $\{x^1,x^2,x^3\}\equiv \{x,y,z\}$,
the corresponding metric components are
$g^\mu_\nu=a \delta^\mu_\nu$. In what follows, a prime will denote
$\diff/\diff\eta$ and an overdot will denote $\diff /\diff t$.

We assume that at early times there is almost exponential inflation,
and set the Hubble parameter $H\equiv\dot a/a$ equal to a constant
$H_*$. Then we can choose $\eta=-(aH_*)^{-1}$. 
After inflation we typically have either radiation domination
when we can choose $\eta=(aH)^{-1}$ or matter domination when
we can choose $\eta=2(aH)^{-1}$. 

\subsection{Spin zero particles}

The formalism in the case of a spin zero field is well known
\cite{physrep,ford,damvil},
 but we reproduce it briefly for comparison with the less 
well-known spin half case. Then we apply it to the case $m(\eta)\sim H$,
which has not been discussed before.

We take as the Lagrangian density of the scalar field $\phi$ 
\be
{\cal L}=\frac12(-{\rm \det\,}g)^{1/2}[\partial_\mu\phi
\partial^\mu\phi-(m^2(\eta)-\xi R)\phi^2 ]
\ee
We have allowed a coupling to the curvature scalar 
$R=-6a''/a^3
=6[\ddot a/a+H^2]$,
but will assume that $|\xi|\lsim 1$.
During exponential inflation, $a\propto=\exp(Ht)$ with $H$ constant
so $R=-12H^2$. During matter domination $a\propto t^{2/3}$ so
$R=-3H^2$, and during radiation domination $a\propto a^{1/2}$
so $R=0$.

With the exception of this coupling, we 
have assumed that all interactions can be represented by
a time dependent mass $m(\eta)$, and as discussed earlier we 
set 
\bea
m(\eta) &=& \mu H(\eta) \mbox{\hspace{10mm}} (\mu H>m) \nonumber\\
m(\eta) &=& m \mbox{\hspace{10mm}} (\mu H<m)
\label{mofeta}
\eea
where $m$ is the true mass and $\mu$ is a constant for order 1.

In order to quantize the theory we 
may start with annihilation operators $a\sk$, which commute except for
$[a_{\bf k},a^\dagger_{\bf k}]=1$. Then $\phi$ is 
Fourier expanded in a comoving box much bigger than the region of 
interest in the form
\be
\phi=a\mone(\eta) \sum_{\bf k}
\left[ w_k(\eta) a_{\bf k}  +w_k^*(\eta) a^\dagger_{-\bf k} \right]
e^{i{\bf k}.{\bf x}}
\label{fourier}
\ee
The equation of motion of $w_k$ is
\be
w_k'' + \Omega_k^2 w_k=0 
\label{kleingord}
\ee
where 
\be
\Omega_k^2\equiv
k^2+ a^2[m^2(\eta)+(\frac16-\xi) R ]
\ee

In the above expansion, the 
factor $1/a$ was pulled out to 
make Eq.~(\ref{kleingord}) simple.
It has the property that the independent solutions $w_k$ and $w_k^*$ 
have a constant Wronskian, which we set equal to $i$,
\be
w_k (w_k')^* -w_k^* w_k' = i
\label{wron}
\ee

If $\Omega_k$ satisfies the 
adiabatic condition
\be
\Omega_k'/\Omega_k \ll \Omega_k
\ee
then $w_k$ is a sinusoidal function of $\eta$ with angular frequency
$\Omega_k$. With our assumptions about $m(\eta)$ and $|\xi|$, 
the adiabatic condition is satisfied whenever 
the expansion rate $a'/a=aH$ is negligible 
compared with $\Omega_k$. This criterion is satisfied
for all all wavenumbers
$k$ in the late-time
regime where $H$ is less than the true mass $m$, whereas at earlier times it
is satisfied only by modes with wavelength inside the horizon,
corresponding to $k\gg aH$,
\bea
k \gg aH &&\mbox{\hspace{10mm}} (H\gsim m) \nonumber\\
{\rm all\,} k&&\mbox{\hspace{10mm}} (H\ll m) 
\label{adiabatic}
\eea
In this adiabatic regime 
we recover
flat spacetime field theory, with $\Omega_k/a$ 
(the physical angular frequency) 
equal to the particle energy 
$E_k=[(k/a)^2+m^2]^{1/2}$.
 Keeping the slow variation
of the adiabatic quantities, the canonical formulation of
flat spacetime field theory corresponds to $w_k=w_k^{\rm {flat}}$,
with
\be
w_k^{\rm {flat}}\equiv\frac1{\sqrt{2aE_k} } e^{-i(E_kt+\chi_k)}
\label{initial}
\ee
where $\chi_k$ is slowly varying and the slowly varying
normalization follows from Eq.~(\ref{wron}).
These expressions lead to the 
particle concept, the number of particles per quantum state
having the expectation value
$n\sk\equiv\langle a\sk^\dagger a\sk\rangle$.

Modes of interest are well inside the horizon now,
and for inflation
to do its job they have to be well inside the horizon at the beginning of 
almost-exponential inflation. (We assume as usual that the
present density is critical, $\Omega_0=1$.)
 Thus they start and finish in the 
adiabatic regime, but if
we choose $w_k=w_k^{\rm flat}$ at early times
we will have 
at late times some solution $\tilde w_k(\eta)$ given by
\be
\tilde w_k(\eta)
\equiv \alpha_k w_k^{\rm {flat}}(\eta)+\beta_k(w_k^{\rm {flat}}(\eta))^*
\label{late}
\ee
From Eq.~(\ref{wron}), $\alpha_k$ and $\beta_k$ are constants 
which satisfy
\be
|\alpha_k|^2-|\beta_k|^2=1
\label{albe}
\ee
The Fourier expansion Eq.~(\ref{fourier}) can be written at late times
\be
\phi=a\mone(\eta) \sum_{\bf k}
\left[ w_k\su{flat}(\eta) \tilde a_{\bf k}  
+w_k^{{\rm flat}*}(\eta) \tilde a^\dagger_{-\bf k} \right]
e^{i{\bf k}.{\bf x}}
\ee
with
where $\tilde a\sk \equiv \alpha_k a\sk + \beta_k^* a_{-\bf k}^\dagger$.
Finally, we suppose that no particles exist at early times,
$\langle a_{\bf k}^\dagger a _{\bf k}\rangle=0$ for all $\bf k$. 
Then the number of particles per state 
at late times is $\langle \tilde a\sk^\dagger \tilde a\sk\rangle
=|\beta_k|^2$.

By solving Eq.~(\ref{kleingord}) with
the initial condition Eq.~(\ref{initial}), we can read off $\beta_k$
from Eq.~(\ref{late}). It is negligible, and hence there is no particle 
creation, if the adiabatic condition Eq.~(\ref{adiabatic}) is satisfied at
all times. This is in general the case only
for modes which are well inside the horizon 
at the end of inflation, ie., for modes with $k\gg a_* H_*$ where a
star denotes the end of inflation. For longer wavelength modes
adiabaticity generally fails leading to significant 
particle creation. 
(Adiabaticity holds for all modes if $m^2(\eta)$
is negligible and $\xi=1/6$, or more generally if
$m^2(\eta)$ and $(\frac16-\xi)R$ cancel, 
but there is no reason to expect these special cases to be realized.)

In the absence of any special reason, one expects that $\alpha_k$ and
$\beta_k$ will be very roughly of the same order, which implies
that very roughly they are both of order 1. One therefore expects
that very roughly the occupation number $|\beta_k|^2$ will be of order 
1 for modes which are inside the horizon at the end of inflation,
and practically zero for shorter  wavelength modes.
In the appendix some specific cases are examined,
and the expectation  verified.

It should be emphasized that we are here talking about the case
where the mass is of order $H$ during inflation. In the opposite case
$m\ll H$ it is well known, from the case of axion physics
\cite{Linde,KT,myaxion}, that 
the occupation number $|\beta_k|^2$ becomes huge in the limit of 
large wavelengths.

\subsection{Spin half particles}

The spin $1/2$ case has been discussed briefly before especially
in the approximation of weak particle production
\cite{parker,spinhalf,dolfreese}, 
but the following account is more systematic and
explicit and we treat for the first time the physically interesting case
$m(\eta)\sim H$.

The starting point is the Dirac equation in 
a Robertson-Walker universe, which can be derived from a Lagrangian 
\cite{spinhalf,dolfreese}
or from more general arguments \cite{parker}. 
With appropriate 
normalization as specified in a moment, it is simply the ordinary 
Dirac equation written in comoving coordinates,
\be
(i a\mone \gamma^\mu\partial_\mu-m +i\frac32 H\gamma^0)\psi=0
\label{dirac1}
\ee
To quantize one can start with the usual
anticommuting annihilation operators 
$a_{r\bf k}$ and $b_{r\bf k}$ (where $r$ denotes the spin state), and make the 
Fourier expansion
\be
\psi=a^{-3/2}\sum_{\bf k}\sum_{\rm r=\pm1}
\left[ u_{r \bf k} a_{r{\bf k}} 
+v_{r \bf k}  b^\dagger _{r-{\bf k}} 
\right]
e^{-i{{\bf k}.{\bf x}}}
\label{dirac}
\ee
(This applies to a Dirac field; for a Majorana field
$a_{r{\bf k}}=b_{r{\bf k}} $ which will make no difference in what 
follows.)
Substituting this expression into the Dirac equation 
and choosing for each mode the $z$ axis along $\bf k$ gives
\bea
i\gamma_0 u'_{r{\bf k}} - [\gamma_z k+am ] u_{r{\bf k}} 
&=& 0\\
i\gamma_0 v'_{r{-\bf k}} - [\gamma_z k+am ] v_{r{-\bf k}} 
&=& 0
\label{dirac2}
\eea
Any solution of this equation can be written as a linear combination
of $u_{1 \bf k} $, $u_{2 \bf k}$,
$v_{1 -\bf k} $  and $v_{2 -\bf k}$.

We impose on these four objects the usual orthonormality condition
\bea
u_{r{\bf k}} ^\dagger u _{s{\bf k}} 
&=& v_{r{\bf k}} ^\dagger v_{s{\bf k}} =\delta_{rs
}
\label{spinornorm1}\\
u^\dagger _{r{\bf k}} v _{s{\bf k}} &=& 0
\label{spinornorm2}
\eea
This condition is preserved by the evolution equation Eq.~(\ref{dirac1}),
because $(\gamma^0)^2$ is the unit matrix
and $\gamma^0( a^{-1}{\bf \gamma}.{\bf k}+m)$ is Hermitian.
Note that we 
did {\em not} pull out a factor $1/(2E_k)$ before defining
$u_{\bf k}$ and $v_{\bf k}$, as is often done in flat spacetime.
Had we done so, $u_{\bf k}$ and $v_{\bf k}$ would have satisfied 
a more complicated 
evolution equation because of the time dependence of $E_k$. We {\em did}
however pull out a factor $a^{-3/2}$, which again is crucial in making
the equation simple.

As the evolution of the two spin states is independent
we need consider only one of them. This
means that we can regard $\psi$ as only a 
two component object so we drop the subscript $r$.
The Dirac matrices can be taken to be
\be
\gamma_0=\left(
\begin{array}{cc} 1 & 0\\ 0 & -1 \end{array}
\right)
\hspace{1cm}
\gamma_z=
\left(
\begin{array}{cc}  0 & 1\\ -1& 0 \end{array}
\right)
\ee

Now drop the subscripts $\bf k$ and $-{\bf k}$, and denote the upper and lower
components of $u$ and $v$ by a subscript $\pm$.
They satisfy uncoupled equations
\bea
u_\pm''+ a^2 [E_k^2\pm i Hm] u_\pm &=&0\nonumber \\
v_\pm''+ a^2 [E_k^2\pm i Hm] v_\pm &=& 0
\label{kg}
\eea
where as before
$E_k$ is the energy related to the momentum
$p_k=k/a$ by $E_k^2=m(\eta)^2+p_k^2$.

With our assumption about $m(\eta)$, 
$Hm\ll E_k^2$  during any era in which the 
adiabatic condition Eq.~(\ref{adiabatic})
is satisfied, and we recover flat spacetime field theory
with the canonical choice
\bea
u^{\rm {flat}}&\equiv&\left(\frac{E_k+m}{2E_k}\right)^{1/2} 
\left(\begin{array}{c} 1\\ p_k/(E_k+m) \end{array} \right)
e^{-i(E _kt +\chi_k)} \nonumber \\
v^{\rm {flat}}&\equiv&\left(\frac{E_k+m}{2E_k}\right)^{1/2} 
\left(\begin{array}{c} -p_k/(E_k+m) \\ 1 
\end{array} \right)
e^{i( E _kt + \chi _k)}
\label{uvflat}
\eea
The ratio of the upper and lower components comes from
Eq.~(\ref{dirac2}), $\chi_k$ is slowly varying,
and the slowly varying normalization comes from 
conditions (\ref{spinornorm1}) and (\ref{spinornorm2}).
As one knows from flat spacetime field theory, these expressions lead to the 
particle concept, the number of particles per quantum state
being $\langle a\sk^\dagger a\sk\rangle$ and the number of antiparticles 
being
$\langle b\sk^\dagger b\sk\rangle$. 

Just as for a scalar field, we consider solutions $\tilde u$ and
$\tilde v$, which become equal to
$u^{\rm {flat}}$ 
and $v^{\rm {flat}}$  well before horizon exit during
inflation. Remembering that $u_+$ has the same evolution equation
as $u_-^*$ and similarly for $v_\pm$, they
are of the form
\bea
\tilde u&=&\left(\begin{array}{c} f_1\\f_2^* 
\end{array} \right)
\nonumber\\
\tilde v&=&\left(\begin{array}{c} -f_2\\f_1^* 
\end{array} \right)
\eea
Thus, in the late time adiabatic regime
they are of the form
\bea
\tilde u(\eta)&=&
\alpha_k u^{\rm {flat} }(\eta)+\beta_k v^{\rm {flat}}(\eta)
\\
\tilde v(\eta)&=&
\alpha_k^* v^{\rm {flat} }(\eta)
-\beta_k^* u^{\rm {flat}}
(\eta)
\eea
Here $\alpha_k$ and $\beta_k$ are constants, which from the 
normalization condition (\ref{spinornorm1})
satisfy
\be
|\alpha_k|^2+|\beta_k|^2=1
\ee
The contribution of the mode to $\psi$ is 
\be
a^{-3/2}\left( u^{\rm {flat} }\tilde a \sk
 + v^{\rm {flat} }\tilde b_{-\bf k}^\dagger
\right)
e^{-i{\bf k}.{\bf x}}
\ee
where 
$\tilde a_\sk = \alpha_k a\sk - \beta_k^* b_{-\bf k}^\dagger$
and $\tilde b_\sk = \beta_k a\sk + \alpha_k^* b_{-\bf k}^\dagger$.
Finally, we suppose that no particles exist at the beginning of 
inflation, $\langle a\sk^\dagger a \sk\rangle=
\langle b\sk^\dagger b \sk
\rangle=0$. Then the number of particles per state 
at late times is $\langle \tilde a\sk^\dagger \tilde a \sk\rangle
=|\beta_k|^2$, and the number of antiparticles
$\langle \tilde b\sk^\dagger \tilde b \sk\rangle $ is the same.

In order to calculate $|\beta_k|^2$ precisely one has to solve the 
second order differential equation (\ref{kg}) with the initial condition
(\ref{initial}). But since it must lie between
0 and 1 one may hope that an adequate order of magnitude estimate 
will be obtained by setting it equal to 1 if there is an era of 
non-adiabaticity and equal to 0 otherwise.
A couple of cases are worked out in the Appendix,
and they support this hope provided that $\mu$ is 
somewhat less than $1$. (For $\mu\sim 1$
one finds $|\beta_k|^2\sim e^{-2\pi\mu}$.)

As in the spin zero case one will have adiabaticity at all times
for those modes which are far inside the horizon at the end 
of inflation. In addition, there is adiabaticity at all times
if the mass is negligible 
in the sense that $p_k(\eta)\gg m(\eta)$ and $p(\eta)^2\gg Hm
(\eta)$, because then Eq.~(\ref{kg})  reduces
to $u_\pm''+k^2 u_\pm=0$.
 However if, as we are supposing, $m(\eta)\sim H$ 
(until the epoch $H\sim m$, which we are supposing occurs after the end 
of inflation), these conditions are equivalent to the adiabatic condition
$p_k(\eta)\gg H$. The 
conclusion is therefore that if $m(\eta)\sim H$ (at least during 
inflation)
one may hope for significant particle production
in all modes for which the adiabatic condition fails just as in the 
spin 0 case. 

\section{Cosmological implications}

We have considered a particle, whose true mass $m$ 
is less than the Hubble 
parameter $H_*$ during inflation. We have assumed that its 
mass $m(\eta)$ in the early universe, before the epoch $H\sim m$,
is of order $H$ and smoothly varying which is what one expects in 
general for a particle with an interaction of only gravitational 
strength. On these assumptions, we have found that
of order 1 particle per quantum state is created for
modes which are outside  the horizon at the end of inflation
(corresponding to $k\lsim a_* H_*$), whereas for smaller scales
particle creation is insignificant. Although we have demonstrated this 
only for spin zero and spin half particles the argument looks quite 
general and it reasonable so suppose that it holds for any 
spin.\footnote
{As mentioned earlier, the gravitino is practically massless during
inflation according to some models but it will have the
canonical mass $\sim H$ afterwards which should be enough to 
ensure that the stated result holds. If the mass were much less than
$H$ even after inflation one would need to ask whether the gravitino
field equation is conformally invariant since in that case
gravitino production would be negligible. We are not sure of the answer 
to that question.}

In this section we see what the estimate implies for cosmology.
We assume of course that the species does not thermalize 
subsequently, since then it loses all memory of its earlier 
abundance.

A convenient measure of the abundance is
provided by the ratio $n/s$, where
$n$ is the number density and $s$ is the entropy density.
After inflation is over and reheating has occurred, this
ratio can be taken to be 
constant between the epoch when the particle stops
interacting and the epoch when it decays.
For a species which is initially in thermal equilibrium,
$n/s\sim g_*^{-1}$ where $g_*\sim 10$ to $10^3$ is the effective number of 
particle species at the epoch when thermal equilibrium fails.

Summing over all modes, the vacuum fluctuation gives
\be
n=(4\pi^2)^{-1} a^{-3} \int^{a_* H_*}_0|\beta_k|^2 k^2 \diff k 
\sim |\beta_{a^*H^*}|^2 (a_* H_* /a)^3
\label{neq}
\ee
In estimating the integral we have assumed that it is dominated
by the upper limit. This is true provided that
$|\beta_k|^2\sim k^\gamma$ with $\gamma>-3$, which holds
for the cases we evaluated in the last section.
As discussed in the text we expect $|\beta_{a^* H^*}|\sim 1$
and from now on we adopt the value 1.

The energy density just after reheating is of order $gT_R^4$ 
where $T_R$ is the reheat 
temperature, and the entropy density is of order $g T_R^3$.
Suppose first that reheating is prompt so that 
$gT_R^4\sim V_*$ where $V_*=3H_*^2 \mpl^2$ is the potential at the end 
of inflation. (Here
$\mpl=(8\pi G)^{-1/2}=2\times 10^{18}\GeV$ is the 
reduced Planck mass.)
Then
\be n/s\sim (H_*/\mpl)^{3/2}\sim (V_*^{1/4}/\mpl)^3
\ee
If reheating is delayed this estimate is reduced by a factor
$T_R/V^{1/4}$ to become
\be
n/s\sim \frac{T_R}{\mpl}\left(\frac{V_*^{1/4}}{\mpl}\right)
^2
\label{vacfluc}
\ee

Now let us compare this abundance with that coming from 
particle collisions, assuming that
the interaction of the 
species under consideration with other particles
is of only gravitational strength.
Candidates for this case are the gravitino (spin
$3/2$) \cite{susy}, the moduli of superstring theory (spin 0) and their 
superpartners the modulini (spin $1/2$) \cite{Banks,Steinhardt}.
Up to numerical factors and dimensionless couplings, the cross section
for producing the species in a typical relativistic particle collision
process is of order $\mpl^{-2}$, and as a result the abundance 
arising from this mechanism is \cite{gravitino}
\be
n/s=f T_R/\mpl
\label{partprod}
\ee
where
$f$ comes from the numerical 
factors and couplings. 
In the case of a gravitino (with the standard properties), 
$f$ can be estimated in 
a fairly model-independent fashion
\cite{gravitino}
to be of order $10^{-5}$.
It is not quite clear what value of $f$ to expect more generally, for
example for moduli and modulini.

Comparing these two estimates, we see that creation from particle 
collisions dominates creation from the vacuum fluctuation, provided that
$V^{1/4}/\mpl \lsim f^{1/2}$. In order not to create too much cmb 
anisotropy one must have \cite{lyth85} $V^{1/4}\lsim 10^{-2}\mpl$, 
and we see that if the upper bound is attained 
the  vacuum fluctuation mechanism is marginally stronger,
at least for gravitinos.

Now let us consider
the cosmological constraints.
First suppose that the species is unstable. Then it should decay
with only gravitational 
strength or as we have just discussed the abundance from
the vacuum fluctuation cannot be significant.
The decay rate is then $\Gamma=10^{-2}\gamma m^3/\mpl^2$ where
the coefficient $\gamma$ is expected \cite{thermal,ewanthermal}
to be  $\lsim 1$.
The corresponding 
decay time $\Gamma^{-1}$ is typically after the beginning of nucleosynthesis,
and as a result the abundance during nucleosynthesis is constrained to 
be \cite{gravitino}
\be
n/s < 10^{-12}\mbox{\rm \ to \ } 
10^{-15}
\label{nucsyn}
\ee
Taking $n/s\lsim 10^{-15}$,
the constraint on the reheat temperature following from
the abundance due to particle collisions is
\be
T_R\lsim 10^{-15} f^{-1} \mpl 
\ee
Assuming that (standard) gravitinos exist they correspond to 
$f\sim 10^{-5}$ which implies the well known constraint
\cite{gravitino}
\be
T_R\lsim 10^{8}\GeV
\ee
Using the vacuum fluctuation abundance instead
multiplies this result by a factor
$f(\mpl/V^{1/4})^2$, which as we just noted might by marginally bigger 
than 1.

Now consider a
stable particle species. 
The
requirement $\Omega_0<1$, where $\Omega_0$ is the present density 
in units of critical density, gives \cite{KT}
\be
n/s< (1\eV/m)
\label{stable}
\ee
where $m$ is the mass of the species. 
Using the abundance from particle collisions,
this gives
\be
\left(\frac{m}{10\MeV} \right)
\lsim \left(\frac{10^{-5}}{f}\right)\left(\frac{10^{16}\GeV}{T_{\rm R}}
\right)
\ee
Using instead the abundance Eq.~(\ref{vacfluc})
coming from the vacuum fluctuation it gives 
\be
\left(\frac{m}{10^5\GeV}\right)
\lsim \left(\frac{10^{16}\GeV}{V^{1/4}}\right)^2 
\left(\frac{10^8\GeV}{T_R}\right)
\label{semifinal}
\ee
If reheating is instantaneous this becomes
\be
\left(\frac{m}{1\MeV}\right)
\lsim \left(\frac{10^{16}\GeV}{V^{1/4}}\right)^3
\label{final}
\ee

\section{Conclusion}

We have presented a complete and unified formalism for the cases 
spin zero and spin half. It invokes minimal physical assumptions, so 
that for example canonical quantization is required only
in the adiabatic regime, with the classical evolution of the
(Heisenberg representation) operators being the only quantum physics
that is used in the intervening regime. The formalism is more complete 
than earlier discussions for the spin half case, and in both cases it is 
here applied for the first time to the case of an effective mass
$\simeq H$.

The bottom line consists of the 
bounds on the parameters presented in the previous section,
associated with the existence of particles having 
gravitational strength interactions.
Though they are are at best only comparable with similar ones derived
on the assumption that the particles are created through collisions
after reheating, they are of interest in principle, and might be
relevant if the inflation scale is high and production through
collisions is more suppressed than for the gravitino.

\section*{Appendix}

In this appendix some specific cases are worked out.

\subsection*{Spin zero}

For modes which are well inside the horizon at the epoch
$H=m$ (equivalently, for modes which are relativistic at that
epoch), it is enough to solve Eq.~(\ref{kleingord}) with
$m(\eta)=\mu H$. 
It then has the form
\be
w_k'' +\left[k^2+\mu^2 \left(\frac{a'}{a}\right)^2+
(6\xi-1)\frac{a''}{a} \right] w_k = 0
\ee
During exponential inflation, radiation domination and matter 
domination, this is equivalent to a Bessel equation for
$w_k/\eta^{1/2}$, with the order $\nu$ given in Table 1,
and assuming that the universe makes abrupt transitions
between these regimes we can calculate $\beta_k$ by matching 
$w_k$ and its first derivative at each transition.

Let us drop the subscript $\bf k$, and let $w\sub{inf}$
denote $\tilde w$ during inflation. For argument $x\gg1$, the Bessel 
function becomes
\be
H_\nu^{(1)}\simeq \sqrt{2}{\pi z} e^{i(z-\frac12\nu\pi -\frac14\pi)}
\label{besslim1}
\ee
By virtue of our assumptions, 
$\nu_{\rm inf}^2=\frac94-\mu^2-12\xi$ is not much bigger than 1
so the regime of large argument corresponds precisely to the
early-time
adiabatic regime
$k\eta\gg 1$. In this same regime $E_k
=k/a$, so one sees that the solution with the required
early-time behaviour 
Eq.~(\ref{initial}) is
\be
w\sub{inf}=\frac{\sqrt\pi}{2} (-\eta)^{1/2} H^{(1)}_{\nu_{\rm inf}}
(-k\eta) \exp[(2\nu_{\rm inf}+1)\frac\pi4 i]
\label{14}
\ee

We shall assume
that inflation gives way 
quickly to an era of either radiation or matter domination, which persists
until the adiabatic regime is encountered.  Then, the solution $w\sub{era}$
of Eq.~(\ref{kleingord}) which reduces to
Eq.~(\ref{initial}) at late times is
the solution $w\sub{rad}$ which 
has the required form Eq.~(\ref{initial}) at late times is
\be
w\sub{era}=
\frac{\sqrt\pi}{2} \eta^{1/2} H^{(1)}_{\nu_{\rm era}}
(-k\eta) \exp[-(2\nu_{\rm era}+1)\frac\pi4 i]
\ee

We are looking for a solution of the form
\be
\tilde w= \alpha w\sub{era} +\beta w^*\sub{era}
\ee
whose value and first derivative 
matches the inflationary solution.
Remembering the normalization
 Eq.~(\ref{wron}) one has
one finds
\bea
\alpha&=& i(w\sub{inf} w^{*\prime}\sub{era}-w\sub{inf}' w\sub{era}^*)
\nonumber\\
\beta&=& i(w\sub{inf} w'\sub{era}-w\sub{inf}' w\sub{era})
\label{alphabeta}
\eea

For modes with $k\gg a_* H_*$ the evolution is adiabatic and
$|\beta|$ is negligible. Considering for simplicity only the opposite 
case $k\ll a_* H_*$, we can use the small small-argument limits of the
Bessel functions,
\begin{eqnarray}
H^{(1)}_{\nu}(x) &=& \frac{-i}{\sin \nu\pi}\left(\frac{1}{\Gamma(1-\nu)
}
\left(\frac{x}{2}\right)^{-\nu} - \frac{e^{-i \nu \pi}}{\Gamma(1+\nu
)}
\left(\frac{x}{2}\right)^{\nu}\right)\\
H^{(2)}_{\nu}(x) &=& \frac{+i}{\sin \nu\pi}\left(\frac{1}{\Gamma(1-\nu)
}
\left(\frac{x}{2}\right)^{-\nu} - \frac{e^{+i \nu \pi}}{\Gamma(1+\nu
)}
\left(\frac{x}{2}\right)^{\nu}\right)
\label{besslim}
\end{eqnarray}
Using them one finds 
\be
\beta=\frac{i}{4\pi} \left(X(\nu_{{\rm inf}},\nu_{{\rm era}}) +
X(-\nu_{{\rm inf}},-\nu_{{\rm era}}) + X(\nu_{{\rm inf}},-\nu_{{\rm era}}) 
+ X(-\nu_{{\rm inf}},\nu_{{\rm era}}) \right)
\end{equation}
where
\begin{equation}
X(a,b) = (1+a+b) e^{i(b-a)\frac{\pi}{2}} \Gamma(-a)\Gamma(-b)
\left(\frac{p\eta_{2}}{2}\right)^{a+b}
\end{equation}

\subsubsection*{Radiation domination after inflation}

For the case of a scalar field with conformal coupling ($\xi = \frac{1}{6}$)
the equations of motion during radiation domination are identical to those
during inflation (ie $\nu_{\rm inf}=\nu_{\rm rad}$). 
For this situation the equation for
$\beta$ simplifies significantly
\begin{eqnarray}
\beta&=&\frac{i}{4\pi}\Bigg( (1+2\nu)\Gamma(-\nu)^{2}
\left(\frac12 \frac k{a_* H_*} \right)^{2\nu} + (1-2\nu)\Gamma(\nu)^{2}
\left(\frac12 \frac k{a_* H_*}\right)^{-2\nu} -\nonumber\\ 
& & {} - {} \frac{2\pi}{\nu}\frac{\cos \pi \nu}{\sin \pi \nu}\Bigg)
\end{eqnarray}
For the case where $c=\mu$ we have 
$\nu= i\sqrt{\frac{3}{4}}$ and hence the third
term dominates
\begin{eqnarray}
\beta& \simeq & \frac{i \cosh \sqrt{\frac{3}{4}}\pi}{\sqrt{3} \sinh
\sqrt{\frac{3}{4}} \pi} \\
|\beta|^{2} &\approx& \frac{1}{3}
\end{eqnarray}
In general the other two (neglected) terms in the formula for $\beta$ will
produce oscillatory terms in the particle numbers, but do not affect the total
number significantly.

For the case of a scalar field without coupling to the Ricci scalar $\nu$ will
in general be different during the different era's under consideration. During
inflation $\nu = \sqrt{\frac{9}{4}-\mu^{2}}$ and during radiation domination
$\nu =\sqrt{\frac{1}{4}-\mu^{2}}$. For $\frac{1}{2} < \mu < \frac{3}{2}$ then
$\nu_{\rm inf}$ will be real whilst $\nu_{\rm rad}$ 
will be imaginary. In this case one of
the terms in the expansion will dominate. If we
set $\nu_{\rm inf} = a$ and $\nu_{\rm rad}= ib$ for 
convenience, where both $a$ and $b$ are
real, then taking the leading term in our expression gives
\begin{eqnarray}
\beta &\simeq&\frac{i}{4 \pi}(1-a-ib)e^{1a\frac{\pi}{2}}e^{b\frac{\pi}{
2}} \Gamma(a)\Gamma(ib)\left(\frac12 \frac k{a_* H_*}\right)^{-a-ib}\\
|\beta|^{2} &\simeq& \frac{e^{b\pi}}{16\pi b \sinh \pi b}
((1-a)^{2}+b^{2}) \Gamma(a)^{2} \left(\frac12 \frac k{a_* H_*}
\right)^{-2a}
\end{eqnarray}
In the case where $\mu=1$ this works out as
\begin{equation}
|\beta|^{2} \approx 0.03\left(\frac12 \frac k{a_* H_*}\right)^{-\sqrt{5}}
\end{equation}

\subsubsection*{Matter domination after inflation}

If a conformally coupled scalar field re-enters the particle horizon during
during matter domination as opposed to radiation domination as discussed above
then the equations describing its evolution will no longer be the same during
the two epochs and hence the full general solution must be used. Taking a
particular choice of mass such that $\mu = 1$ then the orders of the two bessel
functions during the subsequent epochs will be $\nu_{inf} = i\sqrt{3/4}$ during
inflation and $\nu_{mat}=1\sqrt{15/4}$ during matter domination. For these
values it is clear that the third term in the general expression will dominate,
and hence our approximate solution is
\begin{eqnarray}
\beta& \simeq & \frac{i}{4\pi} (1- 1.07i) e^{1.4 \pi} \Gamma
\left(i\sqrt{3/4}\right)\Gamma\left(i\sqrt{15/4}\right)
\left(\frac{k}{a_*H_*}\right)^{-i 1.07} \\
|\beta|^2 & \simeq & \frac{2.145}{(4\pi)^2}e^{2.8 \pi}|\Gamma
\left(i\sqrt{3/4} \right)|^2|\Gamma\left(i\sqrt{15/4}\right)|^2 \\
 & \approx & 0.3
\end{eqnarray}

For a minimally coupled field, $\xi = 0$, 
the equation for $\beta$ is the same as that for re-entry
during radiation domination with the exception that the constant $b$ is
replaced by the corresponding value for matter domination (ie $\nu_{mat} = ib$).
Calculating the result for the case where $\mu = 1$ gives
\begin{equation}
|\beta|^2 \approx 0.03 \left(\frac{k}{a_*H_*}\right)^{-\sqrt{5}}
\end{equation}

\subsection*{Spin half}

In contrast with the spin zero case, 
we definitely need to consider only modes which are
well inside the horizon at the epoch
$H=m$, corresponding to $k\gg a_*H_*$.
 The reason is that modes with smaller wavenumber are 
killed by the 
phase space factor $k^2$ in Eq.~(\ref{neq}),
owing to the fact that the occupation number can 
$|\beta|$ can never exceed 1.
Thus it is enough to set $m(\eta)=\mu H$ at all times.
Then, during exponential inflation, matter domination and radiation 
domination Eq.~(\ref{kg}) is equivalent to a Bessels equation for
 $u_\pm/\eta^{1/2}$ and $v_\pm/\eta^{1/2}$ with the index given
in Table 1. 
During inflation, the solutions reducing to 
Eq.~(\ref{uvflat}) at early times are
\begin{eqnarray}
f_{{\rm inf}(1)}&=&\frac{i}{2}\sqrt{\pi k |\eta|}e^{\mu\frac{\pi}{2}}
H^{(1)}_{\frac{1}{2}-i\mu}(k |\eta|)\\
f_{{\rm inf} (2)}&=&-\frac{i}{2}\sqrt{\pi k |\eta|}e^{-\mu\frac{\pi}{2}}
H^{(2)}_{\frac{1}{2}-i\mu}(k |\eta|)
\eea

We have used this formalism to estimate the occupation numbers
on the assumption that inflation is promptly followed by either
radiation or matter domination. The solutions during radiation 
domination which reduce to Eq.~(\ref{uvflat}) at 
late times are
\bea
f_{{\rm rad}(1)}=-\frac{i}{2}\sqrt{\pi k \eta}e^{-\mu\frac{\pi}{2}}
H^{(2)}_{\frac{1}{2}-i\mu}(k \eta)\\
f_{{\rm rad} (2)}=\frac{i}{2}\sqrt{\pi k \eta}e^{\mu\frac{\pi}{2}}
H^{(1)}_{\frac{1}{2}-i\mu}(k \eta)
\end{eqnarray}
The solutions for matter domination are the same with $\mu$ replaced by
$2\mu$. 
Matching either set of solutions gives
\begin{eqnarray}
f_{{\rm inf} (1)}= \alpha f_{{\rm era} (1)} - \beta f_{{\rm era} (2)}\\
f^{*}_{{\rm inf} (2)} 
= \alpha f^{*}_{{\rm era} (2)} + \beta f^{*}_{{\rm era} (1)}
\end{eqnarray}
And hence
\begin{equation}
\beta=(f^{*}_{{\rm inf} (2)} f_{{\rm era} (1)} - f_{{\rm inf} (1)} 
f^{*}_{{\rm era} (2)})
\end{equation}
where the right hand side is evaluated at the end of inflation. 

For radiation domination this gives
\be
\beta=\frac1{2\pi} \left[\Gamma(\frac12-i\mu)^2\left(
\frac12\frac{k}{a_* H_*}\right)^{2i\mu} - { \rm c.c.}
\right]
\ee
and for matter domination it gives
\be
\beta=\frac1{2\pi} (B
e^{\frac\pi2\mu} - B^* e^{-\frac\pi2\mu})
\ee
where
\be
B=\Gamma(\frac12+i\mu)\Gamma(\frac12+2i\mu)
\left(\frac{k}{a_*H_*}\right)^{-3i\mu}
\ee

Since $|\Gamma(\frac12\pm iy)|^2=\pi/\cosh(\pi y)$
these expressions give $|\beta|^2<1$ as required.
In the adiabatic regimes $k\gg a_* H_*$ and $\mu\gg 1$
they give the expected result
$|\beta|^2\ll 1$. In the 
regime $k\ll a_* H_*$ and $\mu\sim 1$,
they make $|\beta|^2$ a rapidly oscillating
function of $k$ with a mean value roughly of order $e^{-2\pi\mu}$,
which is roughly of order 1 with our assumed value $\mu\sim 1$.

\begin{center}
TABLE 1. The order of the Bessel function\\
\begin{tabular}{c|c|c}
& Scalars & Spinors\\
\hline
Inflation & $\nu^{2}=\frac{9}{4}-12\xi-\mu^{2}$ & $\nu=\frac{1}{2}-i\mu$\\
Radiation dom. & $\nu^{2}=\frac{1}{4}-\mu^{2}$ & $\nu=\frac{1}{2}-i\mu$\\
Matter dom. & $\nu^{2}=\frac94-12\xi -4\mu^2$ & $\nu=\frac12-2i\mu$ 
\end{tabular}
\end{center}

\newpage

\underline{Acknowledgements}

We are 
indebted to Ewan Stewart for many helpful
discussions about supergravity.
The work is partially supported by grants 
from PPARC and
from the European Commission
under the Human Capital and Mobility programme, contract
No.~CHRX-CT94-0423.

\end{document}